\documentclass[12pt,preprint,apj]{emulateapj}

\shorttitle{On the galactic spin of barred disk galaxies}
\shortauthors{Cervantes-Sodi, Li, Park \& Wang}

\begin{document}

\title{On the galactic spin of barred disk galaxies}
\author{Bernardo Cervantes-Sodi \altaffilmark{1}, Cheng Li \altaffilmark{1}, Changbom Park \altaffilmark{2} and Lixin Wang \altaffilmark{1}}
\altaffiltext{1}{Partner Group of the Max Planck Institute for Astrophysics and Key Laboratory for Research in Galaxies and Cosmology
of Chinese Academy of Sciences, Shanghai Astronomical Observatory, Nandan Road 80, Shanghai 200030, China}
\altaffiltext{2}{Korea Institute for Advanced Study, Dongdaemun-gu, Seoul 130-722, Republic of Korea}
\email{bernardo@shao.ac.cn}

\begin{abstract}
We present a study of the connection between the
galactic spin parameter $\lambda_{d}$ and 
the bar fraction in a volume-limited sample of 10,674 disk
galaxies drawn from the Sloan Digital Sky Survey Data Release 7. 
The galaxies in our sample are visually classified into galaxies
hosting long or short bars, and non-barred galaxies. We find that
the spin distributions of these three classes are statistically different,
with galaxies hosting long bars with the lowest $\lambda_{d}$ values, followed
by non-barred galaxies, while galaxies with short bars present typically high
spin parameters. The bar fraction presents its maximum at
low to intermediate $\lambda_{d}$
values for the case of long bars, while the maximum for short bars is
at high $\lambda_{d}$. This bi-modality is in good agreement with previous
studies finding longer bars hosted by luminous, massive,
red galaxies with low content of cold gas, while short bars are found
in low luminosity, low mass, blue galaxies, usually gas rich. In addition,
the rise and fall of the bar fraction as a function of $\lambda_{d}$,
within the long-bar sample, shown in our results,
can be explained as a result of two competing
factors: the self-gravity of the disk that enhances bar instabilities, and
the support by random motions instead of ordered rotational motion, that
prevents the formation/growth of bars.
\end{abstract}


\section{Introduction}
If we take a look of an image of a galaxy such
as NGC 1300, a special feature that immediately catches
our attention is the prominence of the bar. At the same
time, we can find galaxies of similar mass and overall
morphology with no sign of bar presence. The presence of
such prominent non-axisymmetric features must certainly
shape the structure and secular evolution of the host galaxies
by exerting tidal torques which lead to a redistribution
of the galactic constituents (Friedli \& Benz 1993).

Through analytical and numerical calculations previous studies have
shown that bars produce a redistribution of mass and angular
momentum in the galactic disk (Hohl 1971; Weinberg 1985;
Debattista \& Sellwood 2000; Athanassoula 2002;
Martinez-Valpuesta et al. 2006; Hwang et al. 2013).
Stellar density waves can
be driven by galactic bars leading to the formation of
spiral arms (Lindblad 1960; Toomre 1960; Elmegreen \& Elmegreen 1985)
and ring structures (Schwartz 1981; Buta \& Combes 1996).
Bars can also funnel material toward the galaxy center; this 
is specially the case for the collisional gas component that
can dissipate energy during shocks and flow inwards
(Shlosman, Frank \& Begelman 1989; Friedli \& Benz 1993). This inflow
will produce an accumulation of material in the central region
that can help in the build-up of disk-like bulges or pseudo-bulges
(Sheth et al. 2005; Laurikainen et al. 2007; Okamoto 2012),
most prominently in the case of early-type galaxies where bars appear
larger (Elmegreen \& Elmegreen 1985, 1989; Erwin 2005). As a result of
these gas inflows, higher molecular gas concentrations are found in
the central regions of barred galaxies (Sakamoto et al. 1999; Sheth et al. 2005),
as well as younger populations in the bulges of barred galaxies when
compared with unbarred ones (Coelho \& Gadotti 2011).
Numerical simulations by Friedli, Benz \& Kennicutt (1994) show not only
inflow, but also outflow induced by the bar, that can facilitate chemical
mixing and trigger star formation as found in observational studies
(Hummel et al. 1990; Huang et al. 1996; Ellison et al. 2011).

If bars funnel gas inwards and most of massive galaxies host super-massive
black holes (SMBHs), a natural consequence would be an enhanced active galactic
nucleus (AGN) activity. Mechanisms such as "bars within bars" (Shlosman et al.
1989, 2000) and nuclear spirals (Martini \& Pogge 1999; M\'arquez et al. 2000)
have been suggested to transport gas to the vicinity of the SMBH, but 
in this respect no consensus has been reached, with some authors
proposing the presence of bars as a triggering factor for AGN activity
(Knapen et al. 2000; Laine et al. 2002; Laurikainen et al. 2004), while
others are more sceptical (Moles et al. 1995; Mulchaey \& Regan 1997; Lee 
et al. 2012b; Oh et al. 2012).

Some numerical simulations (Hasan \& Norman 1990; Norman,
Sellwood \& Hasan 1996; Athanassoula et al. 2005)
show weakening of the bars, and in some
cases dissolution, by a central mass concentration (CMC)
or gravity torques (Bournaud et al. 2005).
In order to explain the large proportion of barred galaxies
observed in different observational surveys,
mechanisms such as the dissolution and reformation of bars
as recurrent transient structures has been proposed
(e.g. Combes 2000; Bournaud \& Combes 2002), as well as the
survival of bars to the presence of a CMC (Shen \& Sellwood
2004; Debattista et al. 2006)

Recently, the relevance of bars in galaxy evolution has motivated
a large number of studies. The fraction of barred galaxies found
in optical images varies depending on the method used to identify
galaxies with bars and on sample selection, with the typical fraction
between 30\% up to 50\% (Barazza et al. 2008; Aguerri et al. 2009; 
Nair \& Abraham 2010; Lee et al. 2012a, henceforth Lee+12). The fraction
appears even higher when near-infrared images are used, where weak
bars are not obscured by dust and hence are more easily identified
(Marinova \& Jogee 2007; Men\'endez-Delmestre et al. 2007), appearing
at NIR wavelengths as strong bars (Buta et al. 2010). Analyses
of the bar fraction in the Hubble Deep Fields found a dramatic drop
of barred systems at $z > 5$ (Abraham et al. 1999; van den Bergh et
al. 2002) later confirmed by Sheth et al. (2008) studying a sample
of luminous face-on spirals from the 2 deg$^{2}$ Cosmic Evolution Survey,
although results from others studies report the contrary, a constant
bar fraction since $z \le 1$ (e. g. Elmegreen et al. 2004; Jogee et al.
2004). Sheth et al. (2008) also find that the bar fraction in spiral galaxies is a strong
function of stellar mass, color and bulge prominence such as it increases
for more massive, luminous redder systems with a slight preference for
bulge-dominated galaxies. Using a smaller sample of \~190 galaxies
from the Coma Cluster, M\'endez-Abreu et al. (2010) found that bars are
hosted by galaxies in a tight range of luminosities 
(-20 $\lesssim M_r \lesssim $ -17) and masses (10$^{9} 
\lesssim M_{*}/M_{\odot} \lesssim$ 10$^{11}$).

Nair \& Abraham (2010) also report this strong
dependence on mass, but given that their sample contain low mass
galaxies, they are able to assess that the bar fraction is bimodal on mass,
with a high fraction at the low mass end, mostly on late-type spirals,
a decrease of the fraction at intermediate mass galaxies (log$(M/M_{\odot} \sim 10.2$),
and increasing again at the high mass end, this time most of the
barred galaxies being early-type spirals. Recently Lee+12 found that
the bar fraction strongly depends on color, with the fraction increasing
significantly as the color becomes redder for the case of long
bars, while the opposite is found for galaxies hosting short bars
(see also Nair \& Abraham 2010; Masters et al. 2011). Regarding the
dimensions of bars, Erwin (2005) found that early-type disk galaxies
host bars a factor of $\sim$2.5 longer than late-type disks, and Gadotti
(2011) shows that longer bars tend to be hosted by galaxies with
prominent bulges. In general,
bars are found to be larger and stronger in early-type galaxies (Elmegreen
\& Elmegreen 1985, 1989; Aguerri et al. 2009; Hoyle et al. 2011).

All these evidence leads to conclude that bars play a major role in
the secular evolution of galaxies (e. g. Kormendy \& Kennicutt 2004;
Laurikainen et al. 2007) and can be sensitive to dynamical interactions
of the host galaxy with its neighbors (Lee+12).
On previous works, some of us have investigated
the role of the galactic spin on shaping the structure and morphology
of present day galaxies. Features such as galaxy color, the thickness
of the disk, bulge prominence, metallicity  and AGN activity are 
strongly dependent on the spin of the systems (e. g. Hernandez \& Cervantes-Sodi
2006; Cervantes-Sodi \& Hernandez 2009; Cervantes-Sodi et al. 2011).
Numerical studies (Efstathiou, Lake \& Negroponte 1982, henceforth ELN;
Mayer \& Wadsley 2004; Governato et al. 2009; Foyle, Courteau \& Thacker 2008)
have shown that haloes with lower spin parameter, more compact and
self-gravitating, are more prone to global instabilities and
the formation of bars. Syer, Mao \& Mo (1999) tried to prove this hypothesis
using a sample of $\sim$ 2,500 galaxies selected randomly from the 
the ESO-Uppsala catalogue (Lauberts 1982), but could not find any
trend pointing to barred galaxies having systematically lower spin.
It must be taken into account that some systematics could be affecting
the result, such that the majority are relatively late types (mainly 
Sb-Sd types).

Later on, Athanassoula (2008) pointed out that the ENL stability criterion was
over-simplistic and could not distinguish bar stable from bar unstable
disks, specially because it does not take into account the disk
velocity dispersion. On a previous study, Athanassoula \& Sellwood
(1986) showed that velocity dispersion has an important influence
on bar stability. In their studies they presented a disk with no halo,
which was stable because of its high-velocity dispersion.

In the present study we want to address this topic, we will investigate
further if the spin parameter has any direct intervention on the
presence of bars in disk galaxies.
The format of the paper is as follows. In Section 2 we describe the
model to infer the spin parameter for disk galaxies in our sample.
The sample is described in Section 3. In Section 4, we present our
general results of the dependence of the bar fraction on the galactic
spin. Discussion regarding our results is presented in Section 5, 
followed by our conclusions in Section 6.

\section{Estimation of the galactic spin parameter}

To account for the spin of disk galaxies we use the
$\lambda$ spin parameter as defined by Peebles (1971),

\begin{equation}
\label{Lamdef}
\lambda = \frac{L \mid E \mid^{1/2}}{G M^{5/2}},
\end{equation}

where $E$, $M$ and $L$ are the total energy, mass and angular momentum of the configuration, 
respectively. In order to estimate this parameter for the galaxies in our sample, we adopt
the model by Mo, Mao \& White (1998) as we did in Hernandez
\& Cervantes-Sodi (2006).
Here we briefly recall the main ingredients of the model. 
In the framework of the CDM scenario for galaxy formation,
primordial density fluctuations give rise to haloes of dark
matter of mass $M$, within which gas condenses and forms
rotationally supported disks of maximum circular velocity
$V_{d}$. The disk is expected to be thin and presenting an
exponential surface density profile

\begin{equation}
\label{Expprof}
\Sigma(r)=\Sigma_{0} e^{-r/R_{d}},
\end{equation} 

with  $\Sigma_{0}$ the central surface density, $R_{d}$ the disk scalelength,
and corresponding mass

\begin{equation}
\label{mdisk}
M_{d}=2 \pi \Sigma_{0} R_{d}^{2},
\end{equation}

which is a fraction $f_{d}$ of the halo mass

\begin{equation}
\label{massfrac}
M_{d}=f_{d}M.
\end{equation}

If we describe the dark matter halo by a truncated singular isothermal sphere of radius $R_{H}=MG/V_{C}^{2}$, $\lambda = \lambda^{'}$, with

\begin{equation}
\label{Bullock}
\lambda^{'}=\frac{L}{\sqrt{2}MV_{c}R_{H}},
\end{equation} 

as defined by Bullock et al. (2001). The dark matter halo is responsible for
establishing a rigorously flat rotation curve along the disk, from where the
angular momentum of the disk is

\begin{equation}
\label{Lb}
L_{d}=2M_{d}V_{c}R_{d}.
\end{equation}

Assuming that the specific angular momentum of the disk is a fraction $j_{d}$ of that of
the halo, we can express equation (\ref{Bullock}) as:

\begin{equation}
\label{GeneralLambda}
\lambda =  \left({\sqrt{2}}\over{G} \right) \left({f_{d}}\over{j_{d}} \right) R_{d}  V_{d}^{2}M_{d}^{-1}
\end{equation}

We define $\lambda_{d}$ as the product $\lambda_{d}=\lambda j_{d}$. In the case
of both components; baryons and dark matter, having the same specific
angular momentum, $\lambda_{d}$ becomes $\lambda$ and we recover the
expression we have used to assess the spin of disk galaxies in previous works
i.e. Hernandez \& Cervantes-Sodi (2006) and Cervantes-Sodi et al. (2012).
The input parameters in equation (\ref{GeneralLambda}) to estimate the spin of disk
galaxies are the disk scalelenght $R_{d}$, the disk mass $M_{d}$, the circular
velocity $V_{d}$ and the disk mass
fraction $f_{d}$. $R_{d}$ is taken from the SDSS $i$-band. As an estimation for
the disk mass we use the stellar mass obtained from the MPA/JHU DR7 VAGC,
which are based on fits to the SDSS five-band
with the model of Bruzual \& Charlot (2003, see
also Kauffmann et al. 2003). Given that
the location of a galaxy in the Tully-Fisher (TF) relation does not depend
on barredness (Courteau et al. 2003; Sheth et al. 2012),
we determine confidently 
$V_{d}$ by the $r$-band Tully-Fisher relation from Pizagno et al.
(2007) for all the late-type galaxies of the sample. 
Finally, following Gnedin et al. (2007), we derive the disk mass fraction
in terms of the stellar surface density using:

\begin{equation}
\label{Gnedin}
f_{d} = f_{0} \left( \frac{M_{*} R_{d}^{-2}}{10^{9.2} M_{\odot} kpc^{-2} }\right)^{p},
\end{equation}

where $p=0.2$, and $f_{0}$ is chosen using the Milky Way as a representative
example. 
In Cervantes-Sodi et al. (2008) we proved the accuracy in our estimation
of $\lambda$ by comparing the estimation using Equation \ref{GeneralLambda} to
values arising from numerical simulations of different authors. The result
was a one-to-one correlation with small dispersion and no bias, leading
to typical errors of less than $30\%$. It has also been used by other
groups finding it appropriate for different kind of studies (e.g. Puech et al.
2007; Gogarten et al. 2010; Mu\~noz-Mateos et al. 2011).

One would expect that galaxies with low spin parameters, being more compact
and self-gravitating, be more prone to global instabilities. A simple criterion
for the instability of a thin exponential stellar disk was proposed by ELN using a set of N-body experiments; namely, bar instability
occurs if

\begin{equation}
\label{ELN}
\epsilon_{c} \equiv \frac{V_{d}}{(GM_{d}/R_{d})^{1/2}} < 1.1,
\end{equation}

which is a measure of the self-gravity of the disk. This value is to be compared
to a self-gravitation exponential disk, with a value $\epsilon_{c}=0.63$.
By combining equations (\ref{GeneralLambda}) and (\ref{ELN}), we get that 

\begin{equation}
\label{epsilonlambda}
\epsilon_{c}^{2}=\frac{\lambda_{d}}{\sqrt{2}f_{d}}.
\end{equation}

In this context we expect to find an increase of the fraction of
barred galaxies as we move from high to low spinning galaxies.

\section{Data}

The sample of galaxies analysed in this work comes from a previous
study by Lee+12. It is a volume limited sample of
galaxies extracted from the SDSS DR7 (Abazajian et al. 2009) with absolute
magnitude in the $r$-band brighter than $M_r \leq -19.5+5{\rm log}h$
within the redshift range $0.02 \leq z \leq 0.05489$. A total of 33,391 galaxies
are identified from the Korea Institute for Advanced Study Value-Added Galaxy Catalogue
(Choi et al. 2010). Given that we will focus our analysis on late-type galaxies,
we divide this main sample into early- and late-type galaxies using the
prescription by Park \& Choi (2005), where galaxies are segregated according to
their morphology in the color versus color gradient and concentration
index space. An additional visual inspection was performed to improve the
accuracy of the morphology classification.

Galaxies are classified by visual inspection of $g+r+i$
combined color images into long-barred, short-barred and non-barred systems.
A galaxy having a bar that is larger than one quarter of the size of its
host galaxy is classified as long-barred, while the opposite case is a short bar.
The visual classification is more robust for face-on galaxies, where we also
avoid internal extinction effects that would provide an underestimation of $V_{d}$
when using our TF relation. To this effect we limit our sample to galaxies
with the minor-to-major axis ratio $b/a>0.6$. Our final sample contains 10,674
late-type galaxies, 3,240 with long bars and 698 with short bars. As described
in Lee++12, the classification shows a good agreement with the classification
performed by Nair \& Abraham (2010) considering the galaxies in common
in the two samples, with our sample showing only a relatively small fraction of
galaxies with short bars when compared with theirs due to a
slightly stricter criterion. Figure 3 of Lee++12 also shows
some examples of late-type galaxy images with long and short bars.

For a more detailed description of the sample,
and comparisons of the classification
with previous works, we refer the reader to Lee+12.

\section{Results}

\subsection{Dependence of the bar fraction on the galactic spin}

\begin{figure}
\begin{tabular}{c}
\includegraphics[width=.475\textwidth]{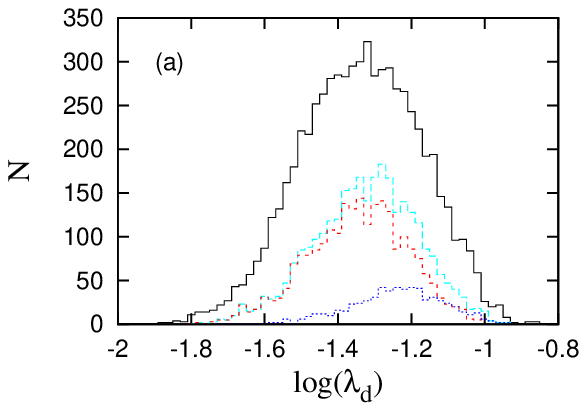} \\
\includegraphics[width=.475\textwidth]{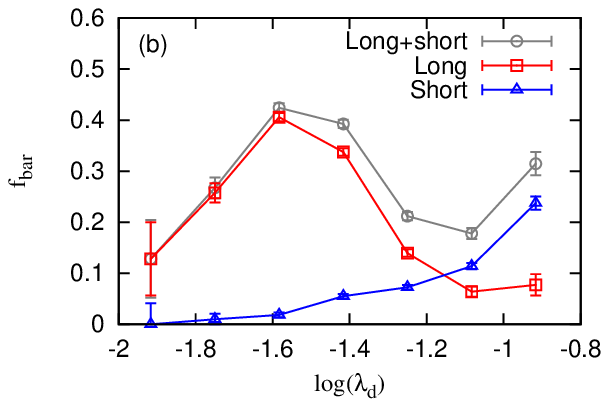} \\
\end{tabular}
\caption[ ]{\textit{(a)}:histograms of $\lambda_{d}$ for non-barred galaxies
(solid black line), barred galaxies
(long plus short bars,long dashed cyan line), galaxies hosting long
(short dashed red line) and short (dotted blue line) bars.
\textit{(b)}: bar fraction as a function of
$\lambda_{d}$ for long, short and long plus short barred galaxies. }\label{SpinDistribution}
\end{figure}

Using our $\lambda_{d}$ estimate from equation \ref{GeneralLambda},
we calculate the spin parameter for all the galaxies in our sample and look
for the empirical distributions $P(\lambda_{d})$ of each of the subsamples
divided according to their morphology.

Figure \ref{SpinDistribution}a shows the spin distribution of disk galaxies in the sample divided in non-barred galaxies, barred ones,
and those hosting long and weak bars. Theoretical (Shaw et al. 2006) and
empirical (Hernandez et al. 2007) distributions of $\lambda_{d}$ are
traditionally described by a log-normal function of the form

\begin{equation}
\label{Plam}
P(\lambda_{d_{0}},\sigma_{\lambda_{d}};\lambda_{d}) d\lambda_{d}=
\frac{1}{\sigma_{\lambda_{d}}\sqrt{2\pi}}exp\left[-\frac{ln^{2}(\lambda_{d}/\lambda_{d_{0}})}
{2\sigma_{\lambda_{d}}^{2}} \right] \frac{d\lambda_{d}}{\lambda_{d}}
\end{equation}

\begin{table}
\begin{center}
\caption{P($\lambda_{d}$) distributions.}
\begin{tabular}{c c ccc}

	\hline

    Sub-sample & $\lambda_{d_{0}}$    &   $\sigma_{\lambda_{d}}$   \\ 
	\hline
	Non-barred & 0.045  & 0.545 \\
	Barred & 0.039  & 0.507 \\
	Long & 0.035  & 0.449 \\
	Short & 0.061  & 0.419 \\	
	\hline
	
\end{tabular}
\end{center}
\end{table}

The parameters describing the different $\lambda_{d}$ distributions
on Figure \ref{SpinDistribution}a are listed on Table 1.
The four distributions are statistically drawn form different underlying
distributions confirmed by Kolmogorov-Smirnov tests, with a significance
level $>$ 99$\%$ for all cases. 
 This shows that long bars are preferentially 
found in low spinning galaxies, while short barred galaxies present high spin
when compared with non-barred galaxies.
In Figure \ref{SpinDistribution}b we present the
bar fraction as a function of $\lambda_{d}$, where we can clearly see two
peaks, one at intermediate-low $\lambda_{d}$ and then an
increase for galaxies with high spin. Error bars in the figure
and subsequent ones denote the
estimated 1$\sigma$ confidence intervals based on
the bootstraping resampling method.

\begin{figure}
\begin{tabular}{c}
\includegraphics[width=.475\textwidth]{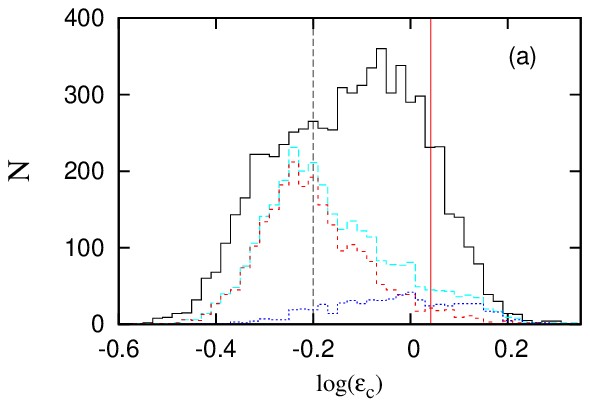} \\
\includegraphics[width=.475\textwidth]{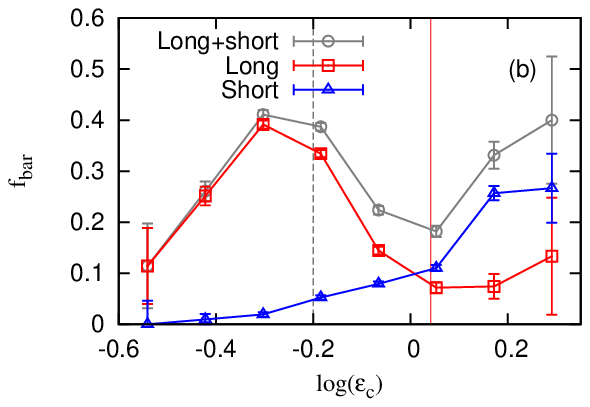} \\
\end{tabular}
\caption[ ]{\textit{(a)}:histograms of $\epsilon_{c}$ for non-barred galaxies
(solid black line), long plus short bars (long dashed cyan line), long
(short dashed red line) and short (dotted blue line).
\textit{(b)}: bar fraction as a function of
$\epsilon_{c}$ for long, short and long plus short barred galaxies. The vertical lines on both figures correspond to the expected value for unstable (solid red line) and self-gravitating (dashed black line) exponential disks.}\label{StabDistribution}
\end{figure}

Dividing again the sample into long and short bars, we identify
the first maximum with the increase of the fraction of long bars, while the
second increase of the bar fraction is due to short bars. As we expected, the bar
fraction is higher for galaxies with low spin parameter,
although this is only for the case of long bars,
the opposite is observed in the case of short bars, with
the bar fraction increasing with increasing $\lambda_{d}$. If we look at the bar fraction
as a function of the disk self-gravity parameter
$\epsilon_{c}$ (Figure ~\ref{StabDistribution}b), we see that only
long bars fulfil the ENL stability criterion, while the frequency of short bars
increases for galaxies that are expected to be stable against bar formation.
It is worth noticing that the bar fraction drops for vanishing spin while still
fulfilling the ENL stability criterion.

\begin{figure*}
\begin{tabular}{cc}
\includegraphics[width=.475\textwidth]{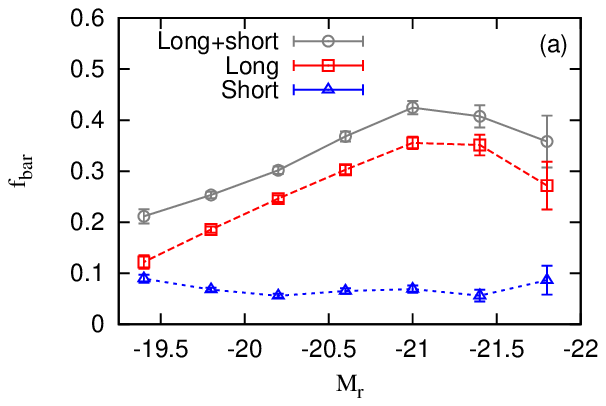} &
\includegraphics[width=.475\textwidth]{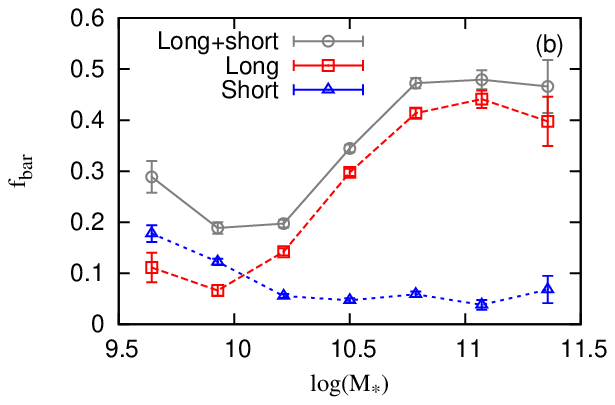} \\
\includegraphics[width=.475\textwidth]{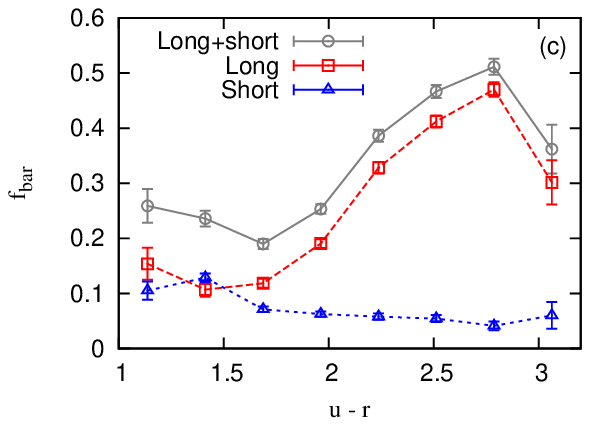} &
\includegraphics[width=.475\textwidth]{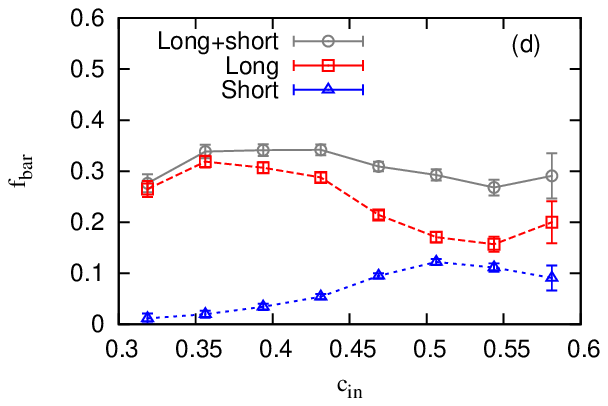} \\

\end{tabular}
\caption[ ]{Dependence of bar fraction as a function of \textit{(a)} $M_{r}$,
\textit{(b)} stellar mass, \textit{(c)} $u-r$ color and \textit{(d)} $c_{in}$.}\label{PrevRes}
\end{figure*}

The difference of the $\epsilon_{c}$-distributions for the different
subsamples shown on Figure~\ref{StabDistribution}a when compared
with the $\lambda_{d}$ distributions on Figure~\ref{SpinDistribution}a,
comes from the dependence on the disk mass fraction as estimated using
equation ~\ref{Gnedin}. Typically, the long barred galaxies have larger
$f_{d}$ values than weak barred or unbarred galaxies. The case of
using a constant dark matter fraction produces a qualitatively similar
result on the dependence of the bar fraction on $\lambda_{d}$, only
enhancing quantitatively the difference on the $\lambda_{d}$-distributions
for the long and short barred galaxies. We present our results using this
more conservative approach.

\subsection{Joint dependence of the bar fraction on the galactic spin and other galaxy properties}

\begin{figure*}
\centering
\begin{tabular}{ccc}
\includegraphics[width=0.30\textwidth]{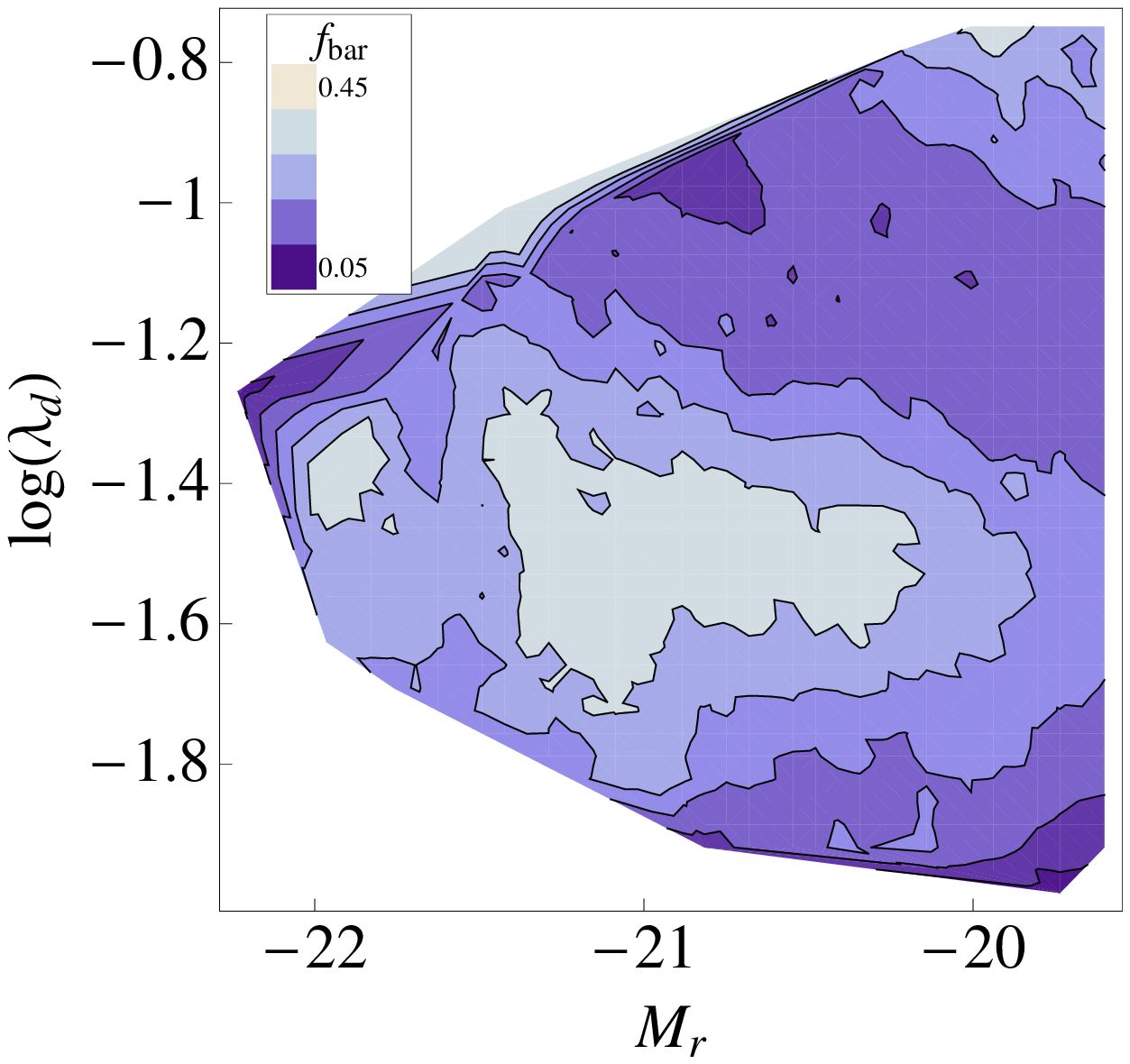} &
\includegraphics[width=0.30\textwidth]{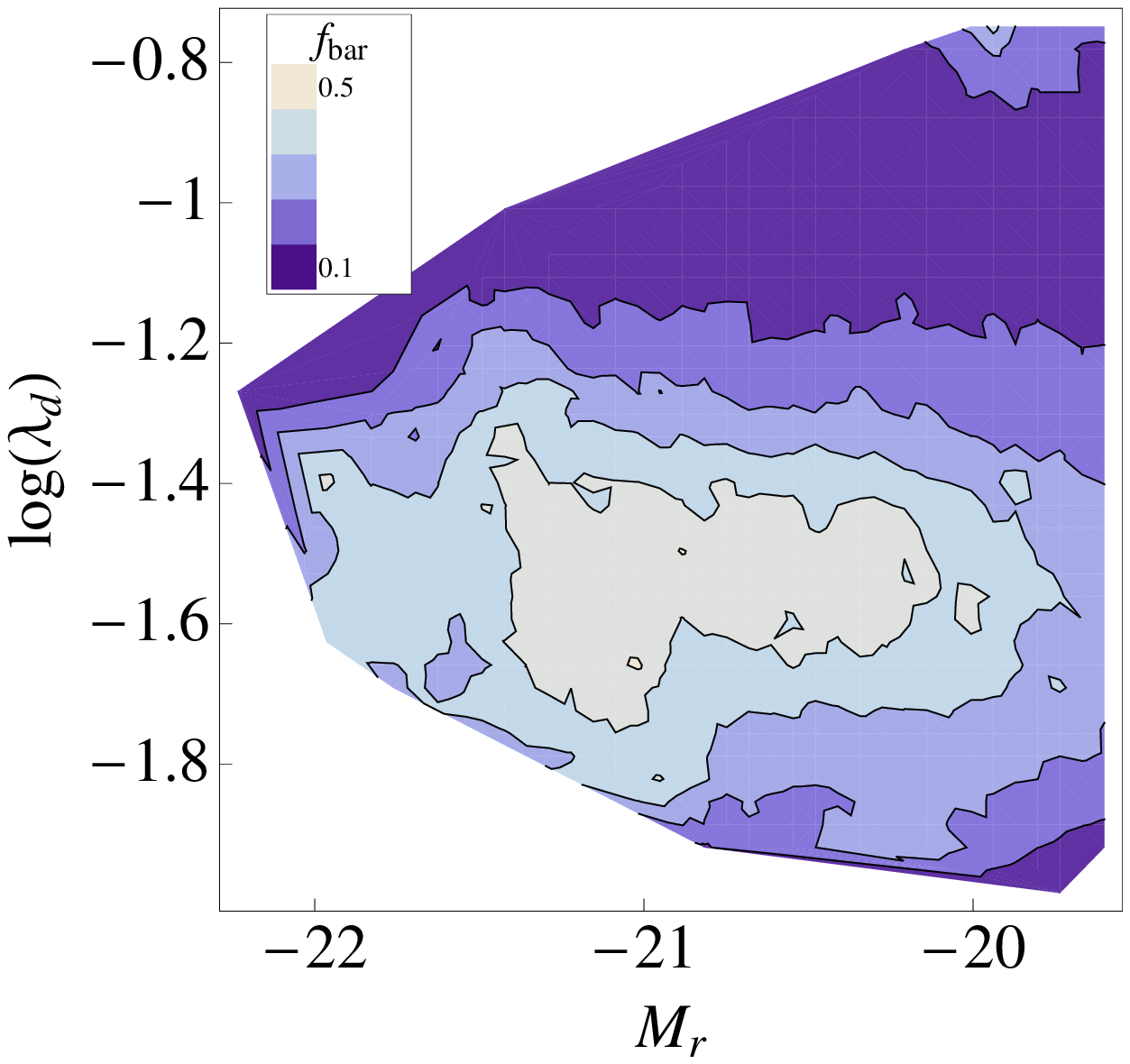} &
\includegraphics[width=0.30\textwidth]{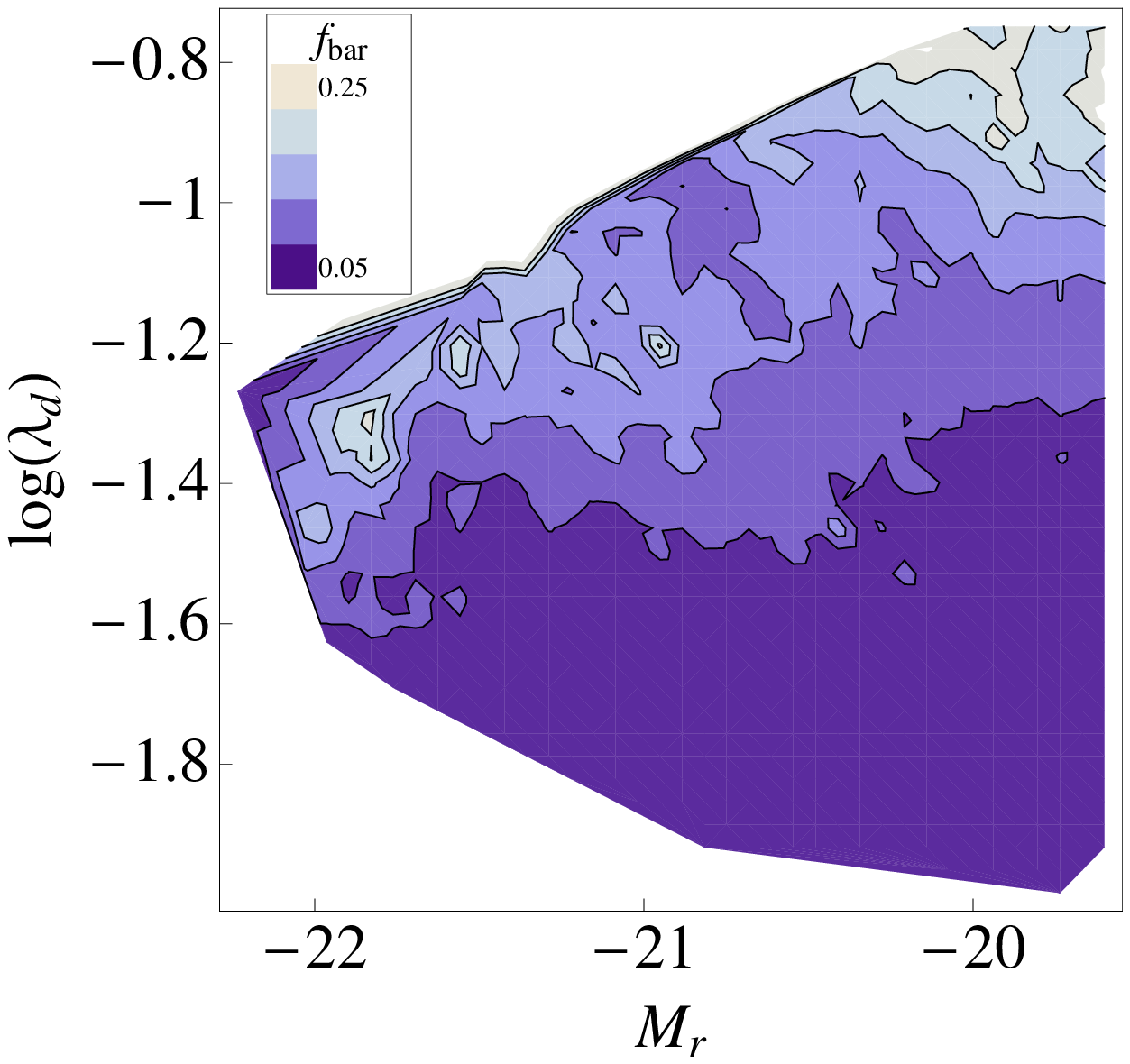} \\
\includegraphics[width=0.30\textwidth]{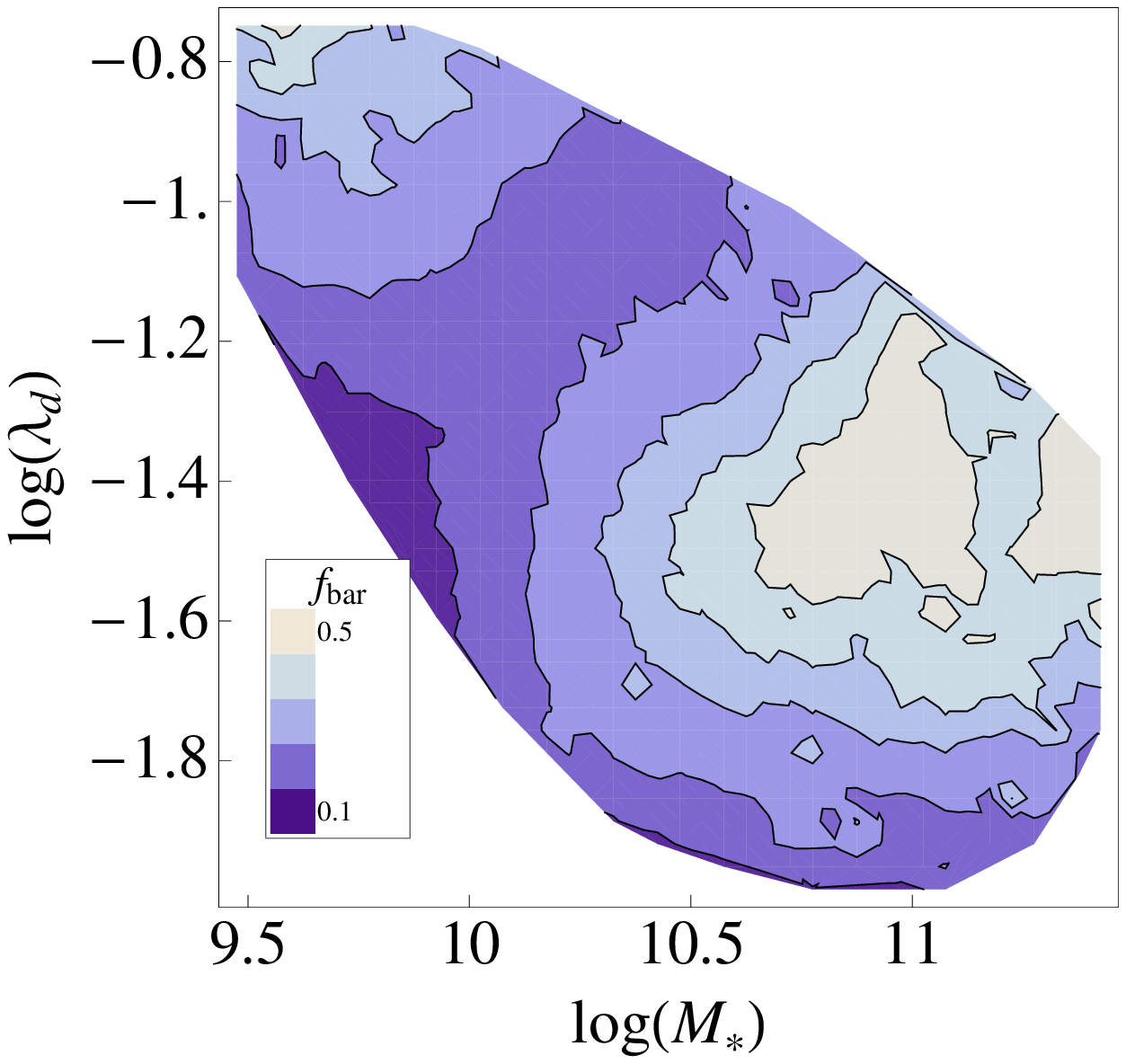} &
\includegraphics[width=0.30\textwidth]{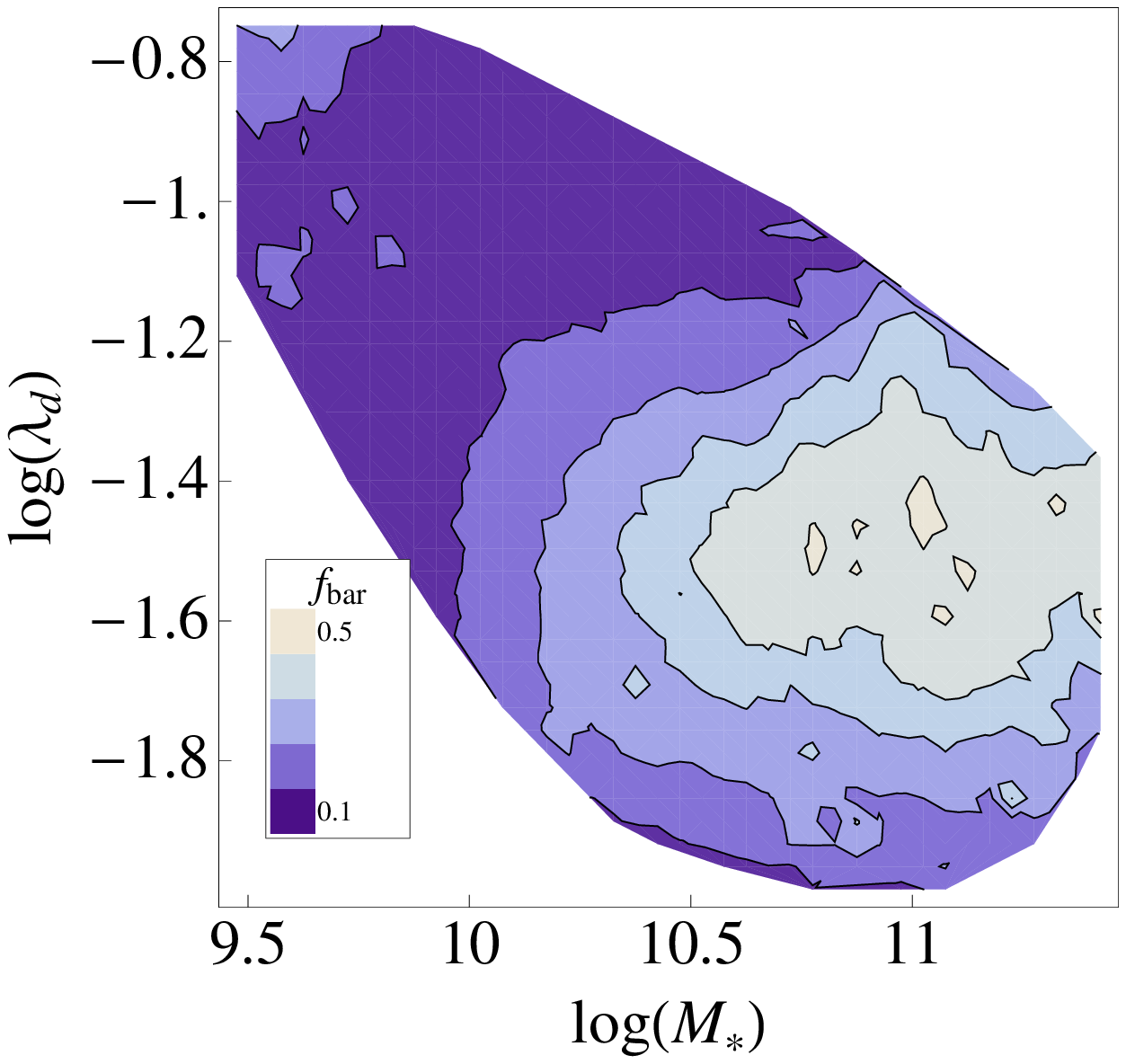} &
\includegraphics[width=0.30\textwidth]{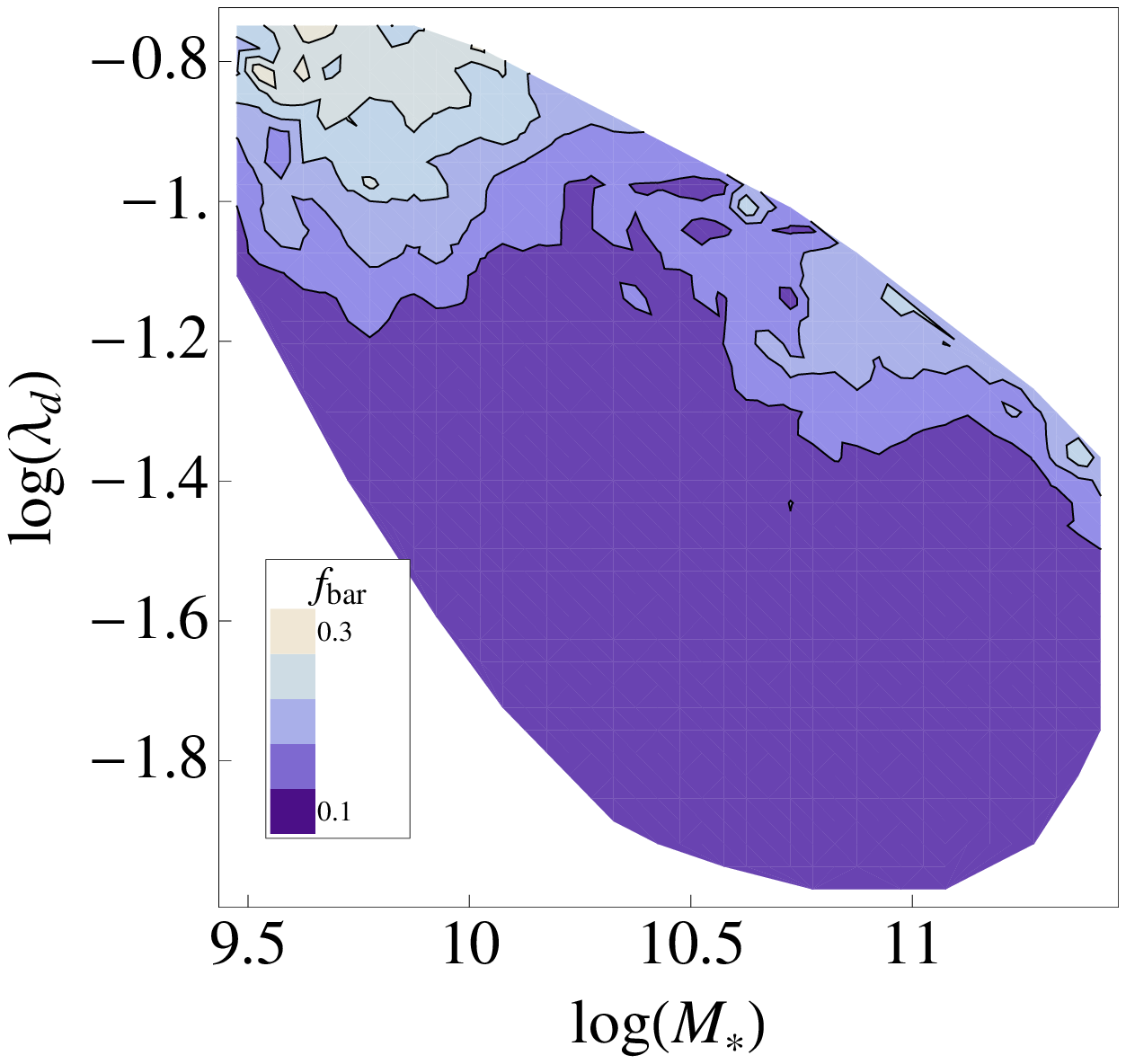} \\
\includegraphics[width=0.30\textwidth]{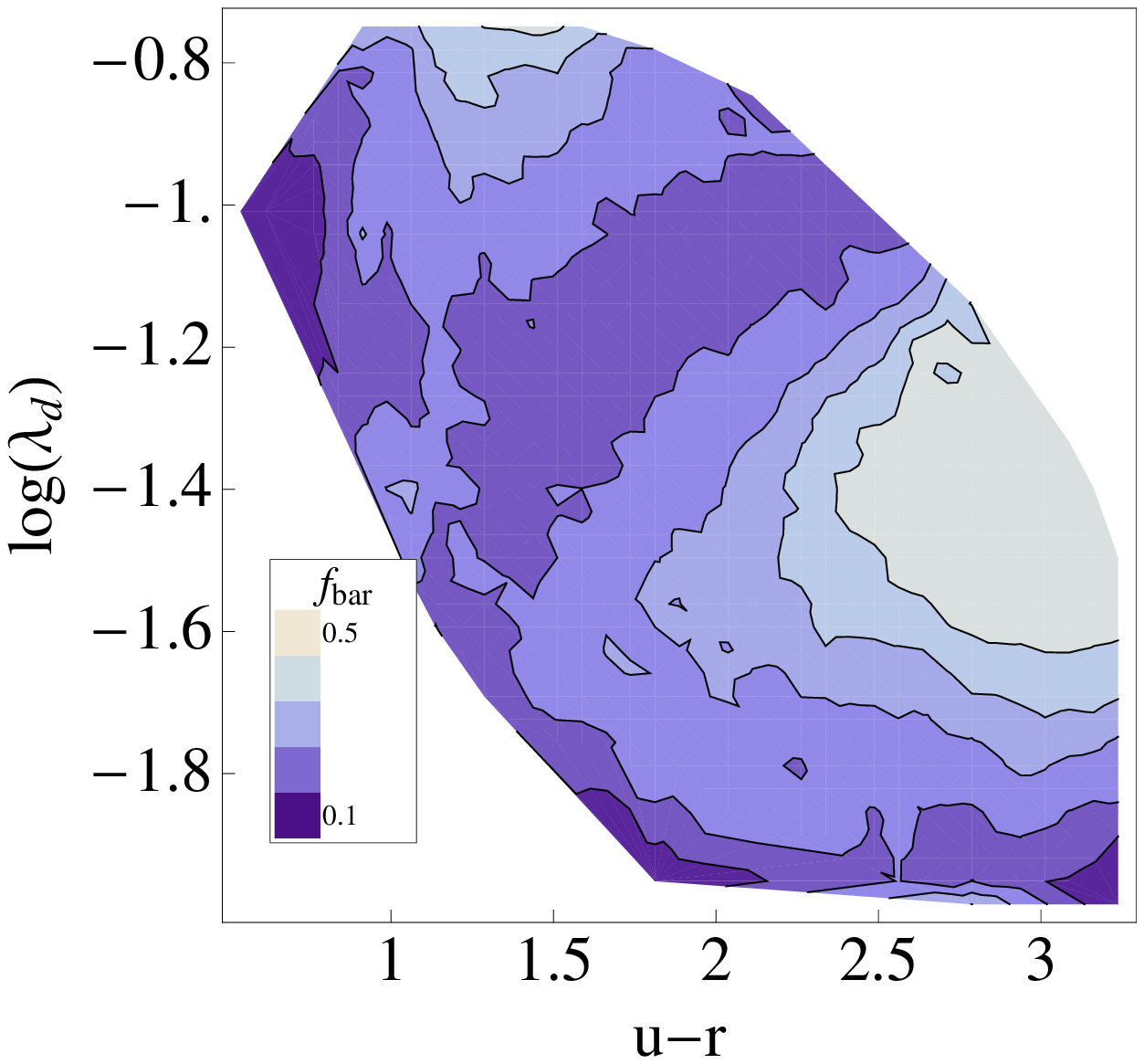} &
\includegraphics[width=0.30\textwidth]{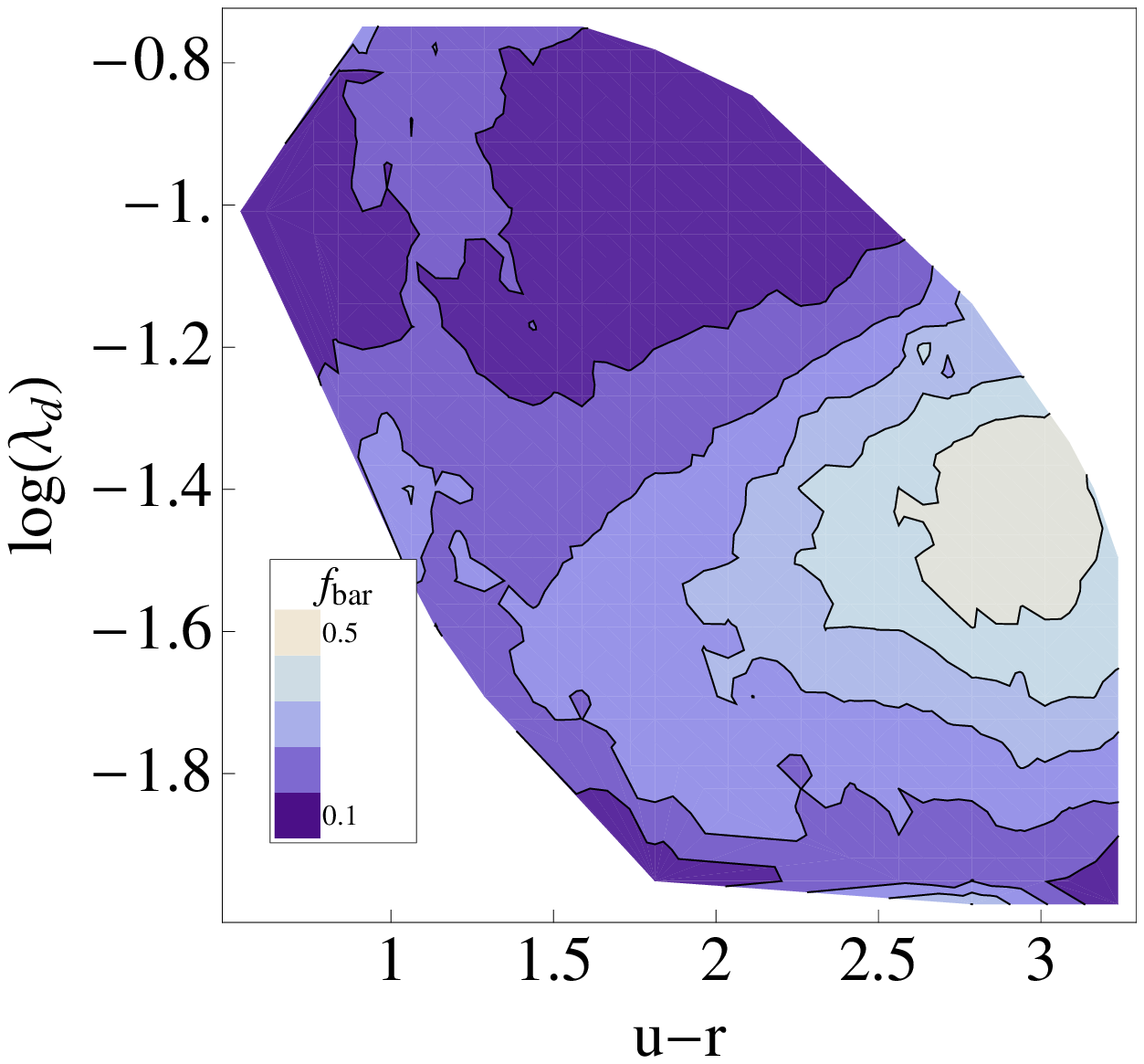} &
\includegraphics[width=0.30\textwidth]{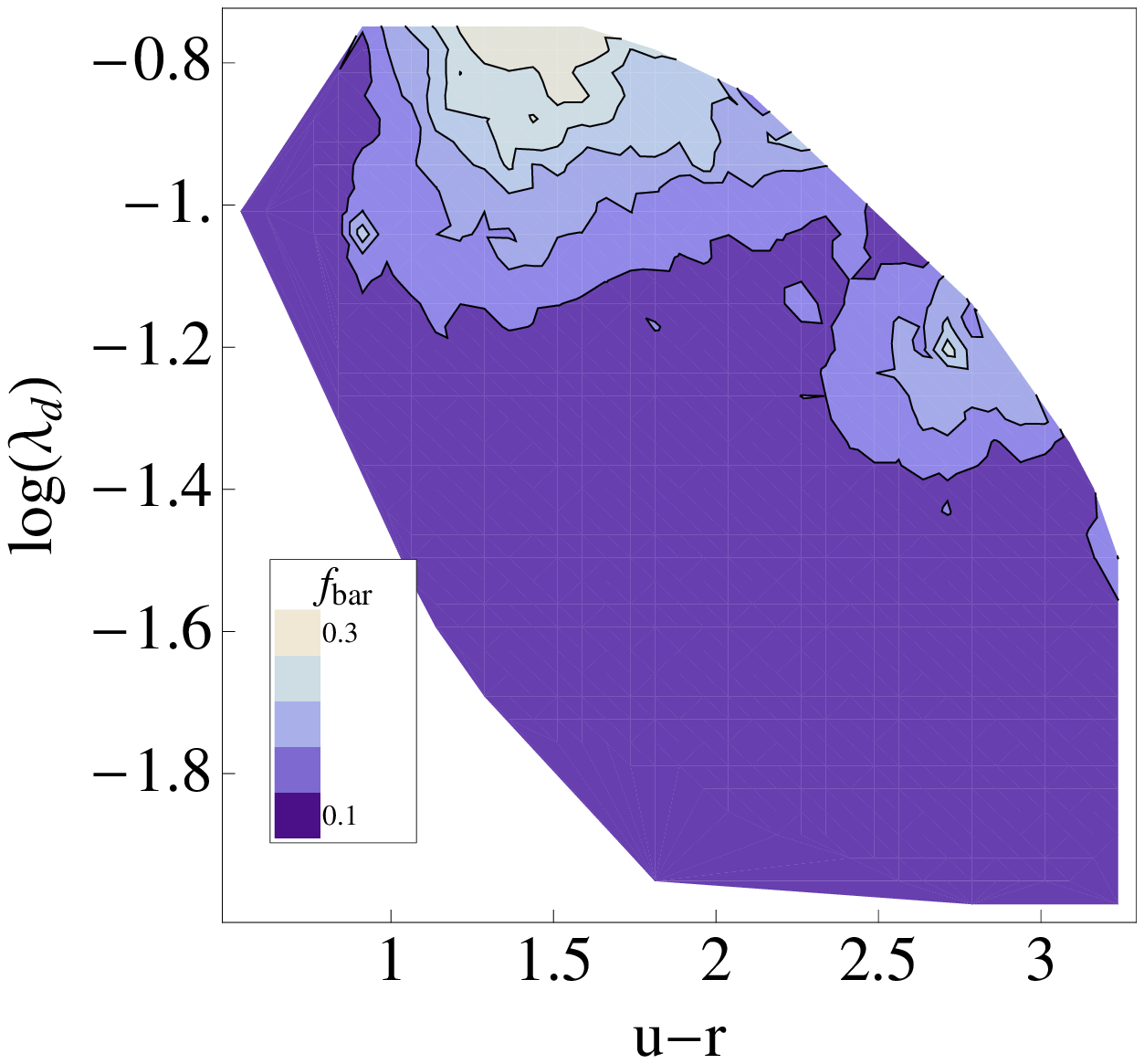} \\
\includegraphics[width=0.30\textwidth]{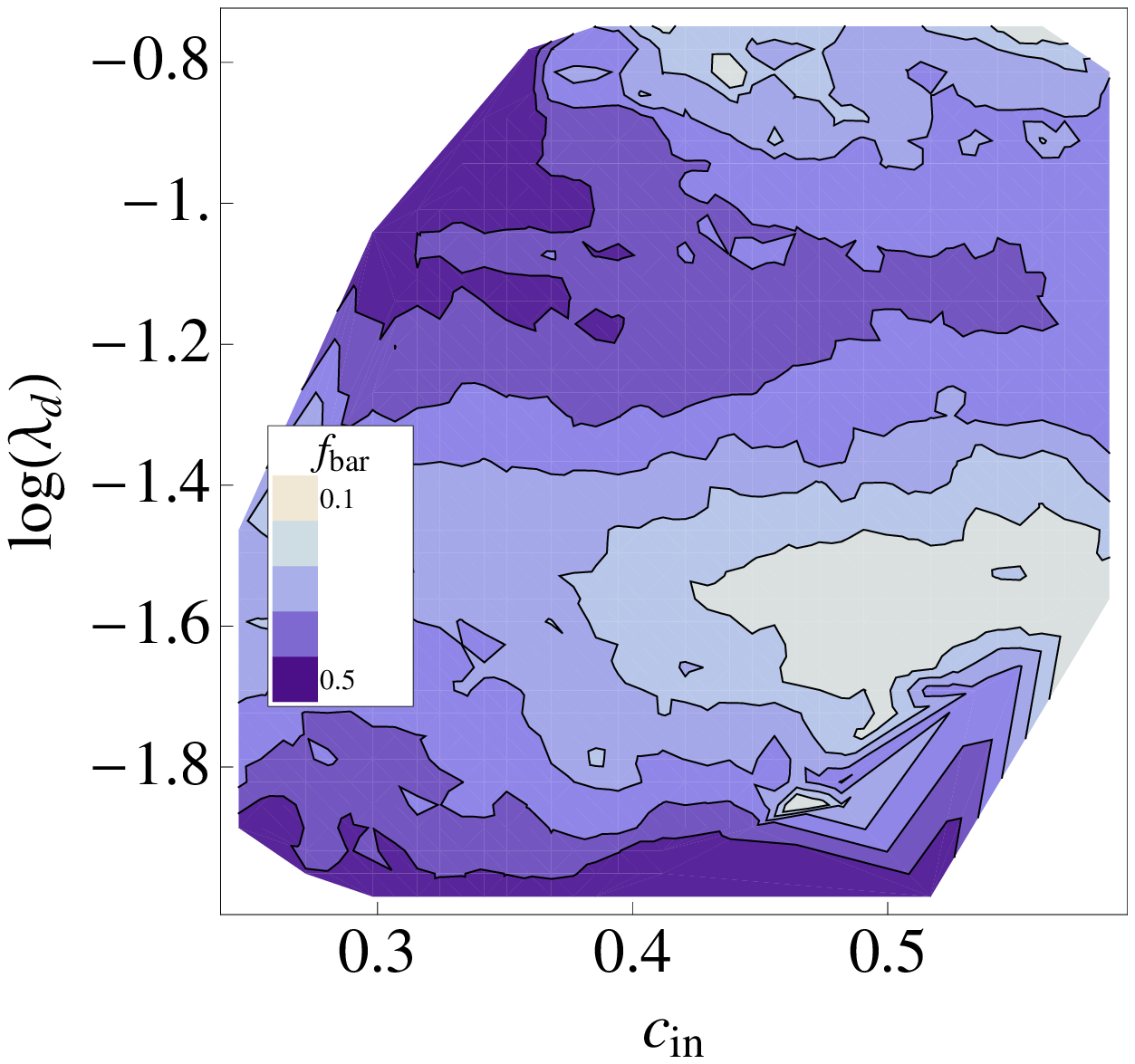} &
\includegraphics[width=0.30\textwidth]{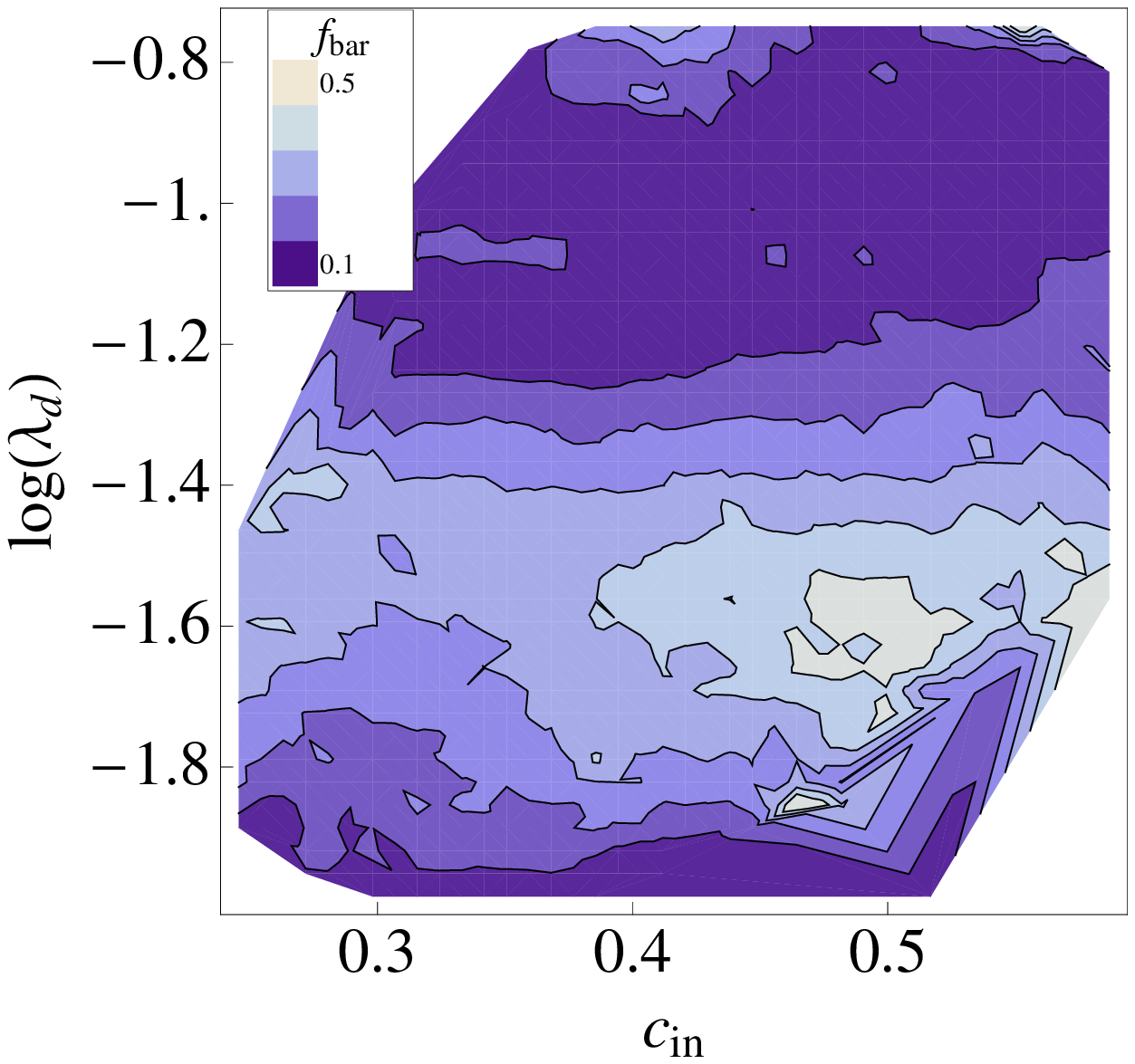} &
\includegraphics[width=0.30\textwidth]{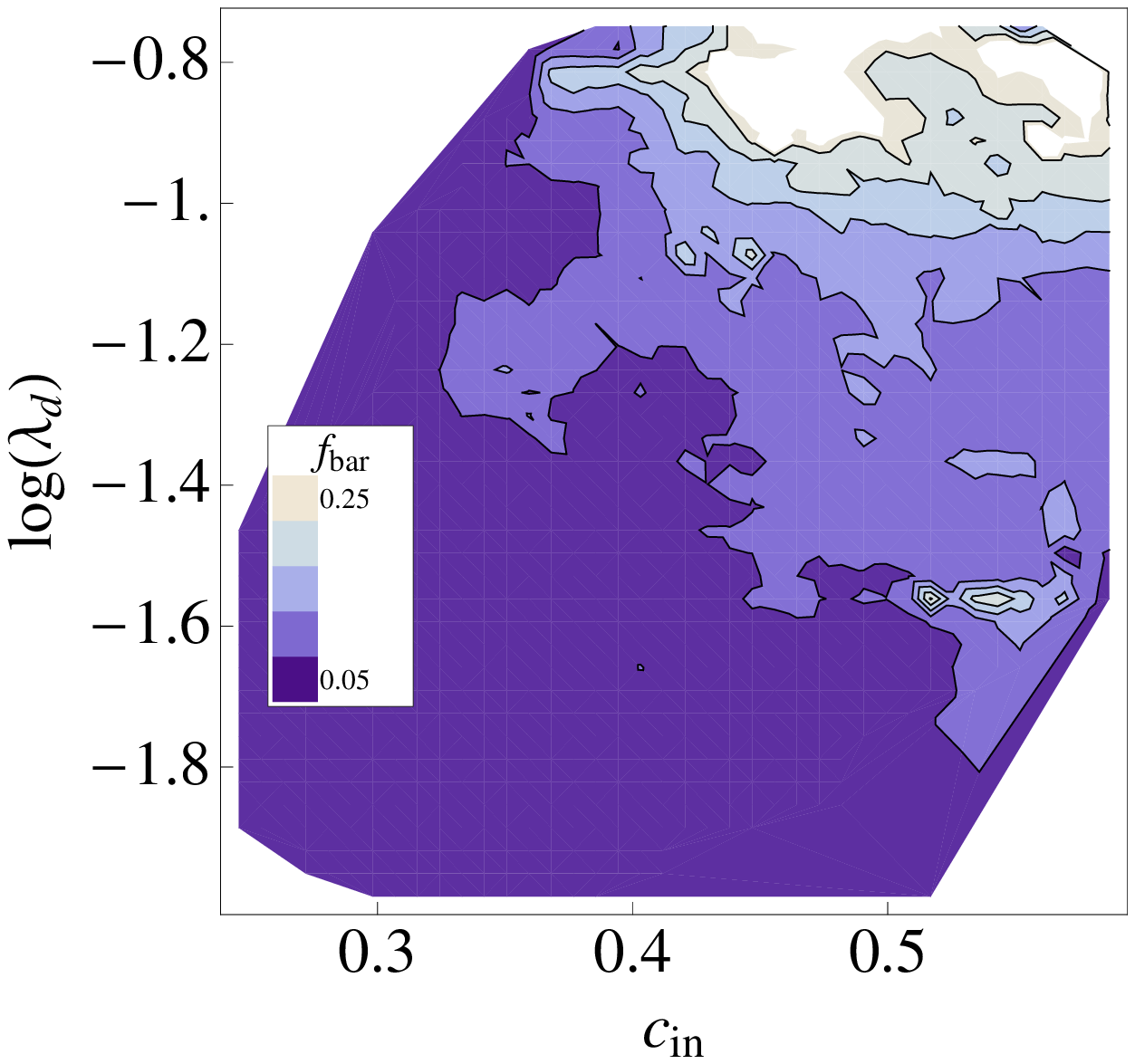} 
\end{tabular}
\caption[ ]{Bar fraction isocontours in the $\lambda_{d}$ vs. $M_{r}$ (\textit{first line}), $M_{stellar}$ (\textit{second line}), $u-r$ (\textit{third line}) and $c_{in}$ (\textit{forth line}) space. Left column correspond to long plus short bars, middle to long bars and right to short bars. Each panel shows their corresponding $f_{bar}$ range and coding.}\label{SpinMaps}
\end{figure*}

Previous works have already shown the dependence of the occurrence of
bars on different galaxy properties, such as luminosity, mass, color, concentration
or gas content. On the other hand, we have also found correlations of the spin
parameter with mass (Cervantes-Sodi et al. 2008), color (Hernandez \& Cervantes-Sodi
2006) and gas mass fraction (Cervantes-Sodi \& Hernandez 2009), among other
parameters. For reference, Figure ~\ref{PrevRes} shows the bar fraction
as a function of absolute magnitude, stellar mass, color and concentration
index, reproducing the findings by Lee+12 and Lee et al (2012b). Long bars
are preferentially found in luminous, massive, red, highly concentrated
galaxies (with low $i$-band inverse concentration
index $c_{in}=R_{50}/R_{90}$), while short
barred galaxies are preferentially blue, less massive galaxies with low concentration indices. Within the magnitude limit of our sample, these
distributions are in agreement with previous findings (i.e. Nair \& 
Abraham 2010; M\'endez-Abreu et al. 2010).

To disentangle the dependence of the bar fraction
on the spin and other crucial physical parameters, we look at the bar fraction in two
parameter spaces involving the spin and other physical properties.
We use a cubic B-spline kernel to get a smooth distribution of $f_{bar}$,
obtaining the ratio of the weighted number of barred galaxies
to the total number of weighted galaxies using a fixed-size smoothing scale .

Figure ~\ref{SpinMaps} top panels presents the bar fraction
in the $\lambda_{d}$ vs. $M_{r}$ space. The first thing to
notice is that even at fixed $M_{r}$ there is a strong 
variation of $f_{bar}$ as a function of the galactic spin for
both cases, short as well as long bars. For the case of long
bars(middle column), its clear that there is a
maximum for bright galaxies with low spin parameters,
although the bar fraction decreases for the
brightest low spinning galaxies of the sample. It is worth noticing
that the contours elongated in the $M_{r}$ direction shows that 
the $f_{bar}$ for long bars shows a stronger dependence on $\lambda_{d}$
than on $M_{r}$. The case of short bars is less clear, at fixed $M_{r}$ there
is a trend of increasing $f_{bar}$ with increasing $\lambda_{d}$, but
no clear trend is found at fixed spin.

The brighter a galaxy is, the more massive it is, specially if we look
at the absolute magnitude in red bands, that better traces the underlying
stellar mass distribution and is less affected by current bursts of star
formation. In Figure \ref{SpinMaps} second line we present the bar
fraction in the $\lambda_{d}$ vs. $M_{*}$ space. $f_{bar}$ for long
bars shows a gradual increase for increasing stellar mass with a maximum
at $11\times10^{10} M_{\odot}$ and then a slight decrease for the most
massive galaxies in our sample. With massive galaxies having in general
low spin parameter, this maximum correspond to low spinning galaxies, but
at fixed stellar mass we can still notice a strong dependence on $\lambda_{d}$,
and actually, for the most massive galaxies in our sample, those with
moderate high $\lambda_{d}$ values are more prone to host long bars.
The symmetry of the contours for the case of long bars tell us that
the dependence on the spin parameter is as relevant as the dependence on
stellar mass. Short bars are mostly present on low mass high spin galaxies

The third line of Figure \ref{SpinMaps} presents the bar fraction in the
$\lambda_{d}$ vs. $u-r$ space. The double dependence on the chosen parameters
is again noticeable. Although in our sample we can find blue galaxies with
low spin parameter that fulfil the stability criterion for the formation of bars
(equation \ref{ELN}),
the long bar fraction of these galaxies is low when compared with red galaxies where
$f_{bar}$ increases. Among red galaxies, systems with high spin present the
highest bar fraction. This is also the case of short bars, but they reside in
blue galaxies.

On the last line of Figure \ref{SpinMaps} we present the bar fraction in
the $\lambda_{d}$ vs. $c_{in}$ space. It is clear in this case that once
we fix a specific value for $\lambda_{d}$, there is very little variation of
the bar fraction varying the concentration index, with the exception of the
less concentrated systems, specially for the case of long bars. 
Given this, we can attribute the dependence of $f_{bar}$ on $c_{in}$ directly
to the spin of the galaxies.

\subsection{Dependence of the bar fraction on random motions}

\begin{figure}
\begin{tabular}{c}
\includegraphics[width=.475\textwidth]{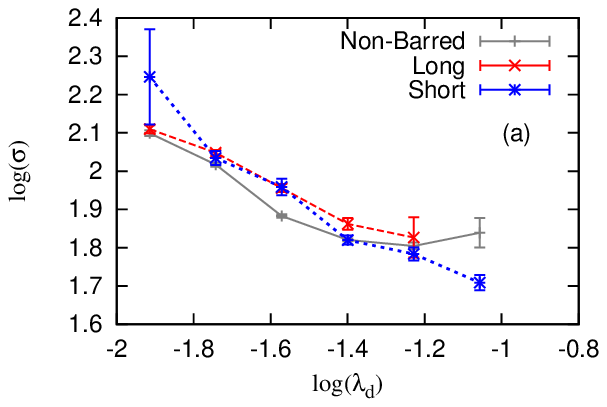} \\
\includegraphics[width=.475\textwidth]{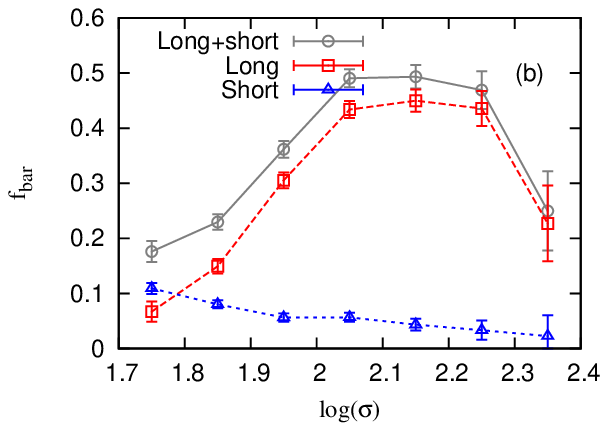} \\
\end{tabular}
\caption[ ]{\textit{(a)}: Dependence of $\sigma$ on $\lambda_{d}$.
\textit{(b)}: bar fraction as a function of
$\sigma$. }\label{SigmaFrac}
\end{figure}

As shown in Figure \ref{SpinDistribution},
the dependence of the bar fraction on the spin parameter is
different for long and short bars.

For the case of galaxies with long bars, the bar fraction reaches
its maximum at intermediate-low $\lambda_{d}$ values. Since it is the self-gravity of
the disk which drives dynamical instabilities responsible for the formation
of the bar, galaxies with low spin, being less dispersed, will be more
prone to host a bar. Our finding of decreasing bar fraction with increasing
$\lambda_{d}$ is in good agreement with the vanishing bar fraction on
galaxies expected to be stable against bar formation as accounted by the
ELN stability criterion.

It is important to point out that the maximum bar fraction is at low
spin values, but as we approach zero, the bar fraction decreases dramatically.
These galaxies are expected to be dynamically unstable and prone to the
formation of bars as they are the most self-gravitating systems, but 
at the same time, as their spin value decreases, their rotational support
is also diminished and random motions are expected to start dominating.
Numerical studies (Athanassoula \& Sellwood 1986; Athanassoula 2008)
show that disks that are expected to be unstable
in terms of the ELN stability criterion become stable against bar formation
due to their high-velocity dispersion, delaying the formation of a bar.
Observational results confirm this hypothesis. Das et al. (2008) found on a
sample of local galaxies that the bar strength and the normalized central
velocity dispersion anti-correlate, suggesting that the bar weakens as the
central component becomes kinematically hotter, and more recently,
Sheth et al. (2012) analysing a sample at higher redshift,
found that bars where not present in dispersion-dominated disk galaxies.
This gives a good explanation to the vanishing bar fraction as
$\lambda_{d}$ approaches zero.

To give a further proof that this mechanism might be playing a role
on the declining fraction of long bars in low spinning galaxies, we
first investigate if the velocity dispersion shows any correlation
with the spin parameter or the bar fraction. To get the minimum contribution
from the rotation curve we limited our sample to those galaxies with $b/a>0.8$.
Figure \ref{SigmaFrac}a shows a clear anti-correlation of the
velocity dispersion with the spin parameter that points in the direction
we expect if we require low spinning galaxies to be supported by random 
motions. As already pointed out by Lee+12, the long bar fraction reaches
its maximum at intermediate velocity dispersion values in the range 
125$-$175 km s$^{-1}$, as we notice in Figure \ref{SigmaFrac}b.
Lee+12 also concluded that the decline of the bar fraction for $\sigma >$
175 km s$^{-1}$ could be due to the disruption of bars in galaxies
with centers dynamically too hot, or that bar formation was
ultimately suppressed on such systems (see Athanassoula et al. 2013).

\begin{figure*}
\centering
\begin{tabular}{ccc}
\includegraphics[width=0.30\textwidth]{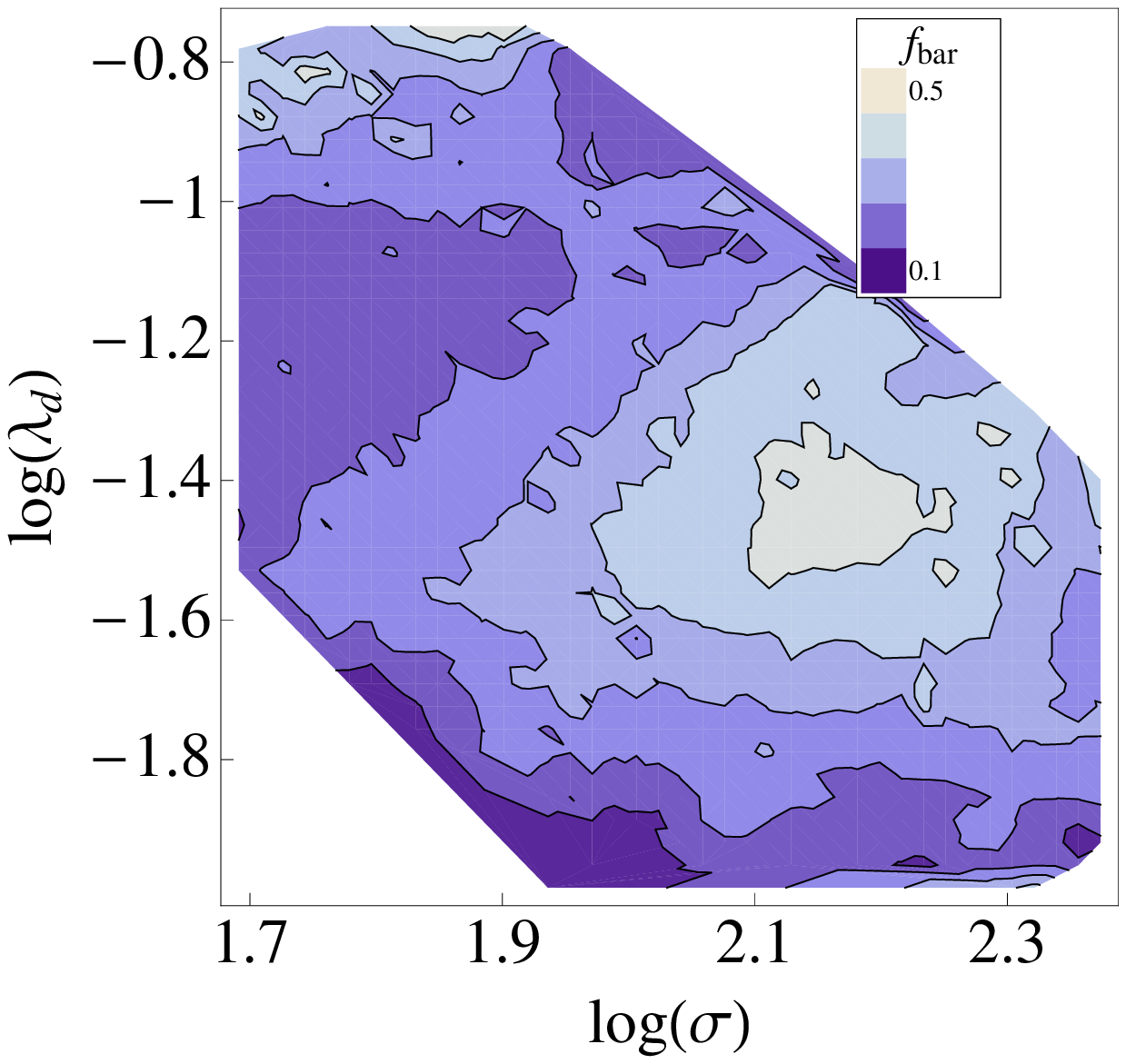} &
\includegraphics[width=0.30\textwidth]{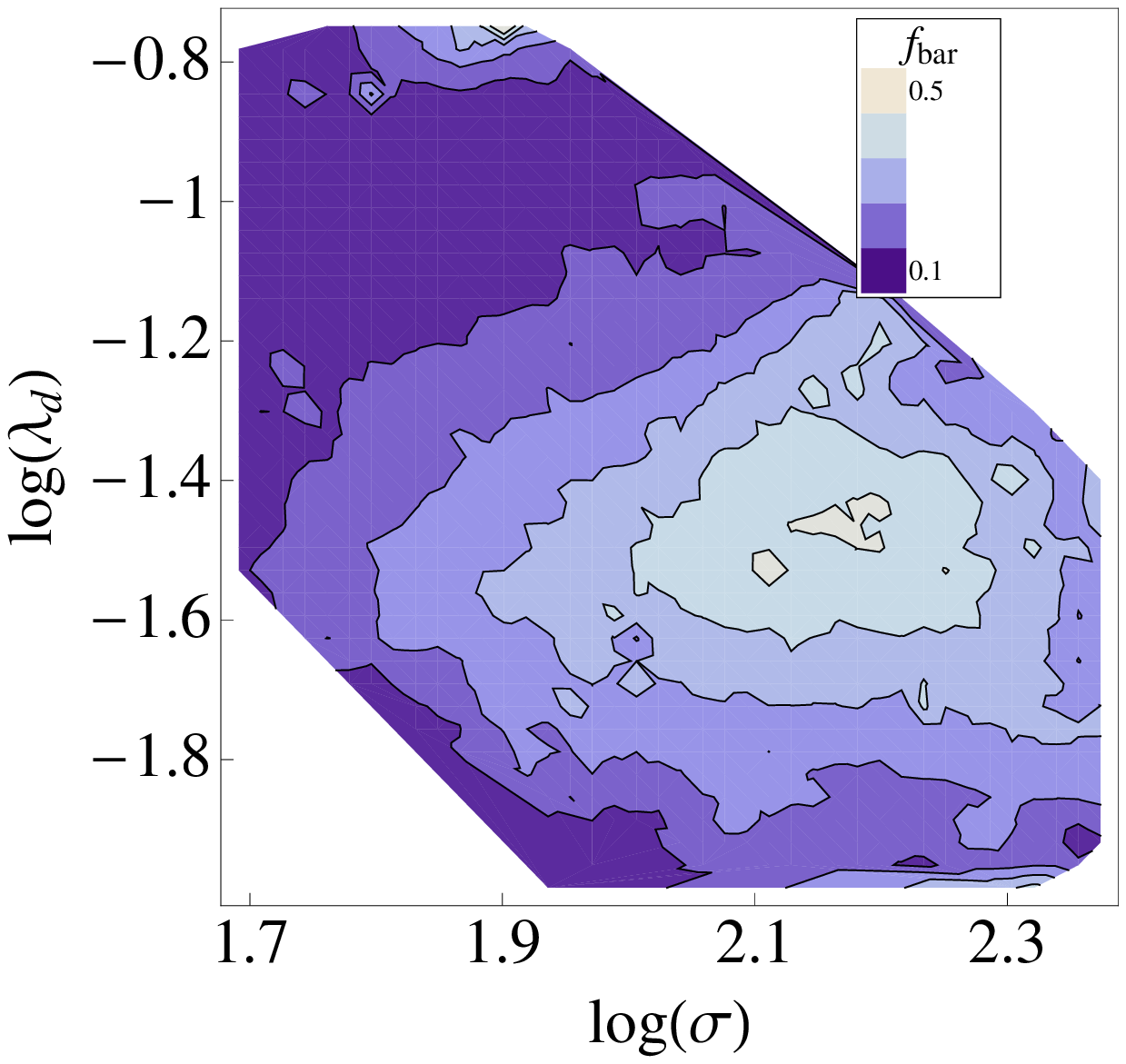} &
\includegraphics[width=0.30\textwidth]{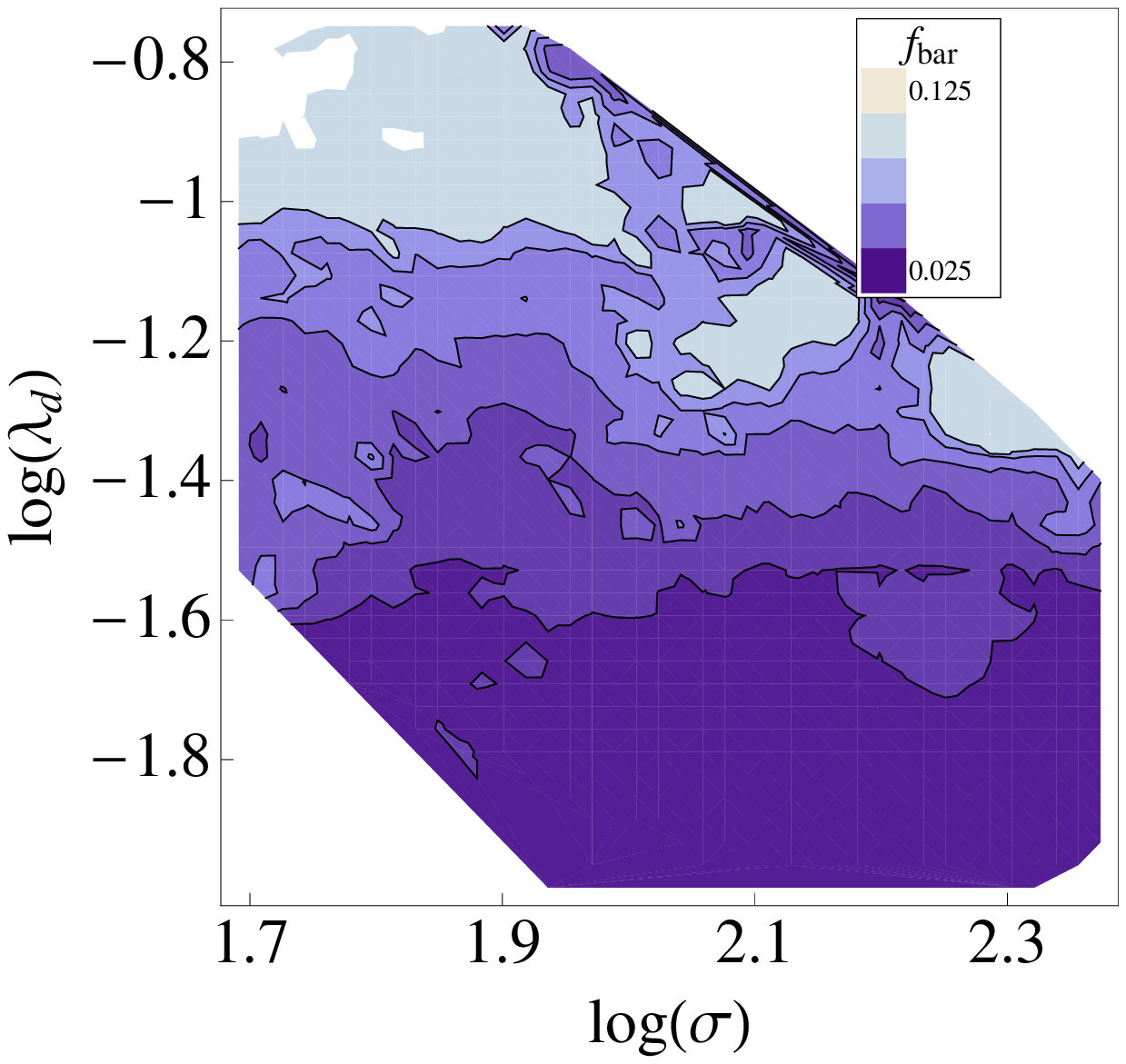} 
\end{tabular}
\caption[ ]{Bar fraction isocontours in the $\lambda_{d}$ vs. $\sigma$ space. Left column correspond to long plus short bars, middle to long bars and right to short bars. Each panel shows their corresponding $f_{bar}$ range and coding. \vspace{1cm}}
\label{SigmaMaps}
\end{figure*}

The corresponding two dimensional maps for the bar fraction in the
log($\sigma$) vs. log($\lambda_{d}$) space are shown in Figure \ref{SigmaMaps}.
In agreement with what we expected, we can first notice the anti-correlation
between the two physical parameters, and for the case of long bars, a 
clear decrease of the bar fraction, even for low spinning systems, with 
$\sigma >$ 160 km s$^{-1}$. Although this cannot be regarded as a prove
for our hypothesis, given that our $\lambda_{d}$ estimation comes from
invoking a TF relation, and not from a pure kinematic study, this result
can be explained as a result of two competing factors, the self-gravity of
the disk that enhances the formation of bars, and the support by random
motions that works in the opposite direction, preventing bar instabilities.

\section{Discussion}

Galaxies hosting
a short bar have typically higher spin than those hosting long bars or non-barred
galaxies, with the bar fraction increasing with increasing $\lambda_{d}$.
The presence of these short bars on high $\lambda_{d}$ systems confront the
ELN stability criterion as shown in Figure ~\ref{StabDistribution}b,
where the bar fraction increases even for galaxies with $\epsilon_{c}>1.1$.
As pointed out by Sellwood \& Moore (1999), while the normal bar instability
gives a natural explanation for the presence of strong bars, there is no
alternative mechanism to explain the existence of weak bars.
Erwin (2005) presents a careful and extensive comparison of
bar lengths between observational and numerical results and concludes that
the simulations compared with his results tend to produce long bars, and
except two early simulations (Pfenniger \& Friedli 1991; Combes \& Elmegreen
1993), no N-body bars are as small as typical Sc-Sd bars, 
that are expected to be galaxies with a high spin parameter.

An extreme case of high spinning systems is low surface brightness galaxies
(LSBs). Simulations with $\lambda_{d}>0.04$ are able to reproduce
some of the physical characteristics of these galaxies such
as their surface density profiles and colors (e. g. Jimenez et al. 1998;
Kim \& Lee 2012), and while they are expected to be stable to bar
formation due to the low self-gravity of their disks and be
dark matter dominated at all radii, some simulations are able to produce bars that
display a mild oval distortion (Mihos, McGaugh \& de Block 1997), with
short-lived episodes that finally evolve to a bulge-like structure (Mayer \&
Wadsley 2004). 

The fact that bars appear short on high spinning galaxies could be due
to the low efficiency of their disks as angular momentum sinks. The
disperse disks of these systems are unable to carry away the necessary
amount of angular momentum from the bar in order to grow to the extent
of the typical bars of low spinning galaxies.
  
Recently, Scannapieco \& Athanassoula (2012) studied the properties
of bars formed in fully cosmological hydrodynamical simulations of
Milky Way-mass galaxies and found that the longest bar is formed in
a bulge-disk-halo system with the lowest spin parameter, and the shortest is
found in a galaxy with a disk but not significant bulge and the highest 
spin parameter. Previous studies have shown similar results (Athanassoula \&
Misiriotis 2002; Ahtanassoula 2003), stronger bars residing in galaxies with
prominent bulges. Hernandez \& Cervantes-Sodi (2006) noticed that
galaxies with the largest bulge-to-disk ratio are those with low spin
parameter, a result that fits in this picture of galaxies with low
values of $\lambda_{d}$ presenting longer bars that galaxies with high
spin. 

The competing effects of self-gravity in development and growth
of the bar instability and the suppression through random motions in
dispersion dominated systems helps to explain the dependence of the
bar fraction on the galactic spin. Given the clear trend for mean $\lambda_{d}$
values increasing in going towards later Hubble types (Cervantes-Sodi \& 
Hernandez 2009 and references therein), our results are in good agreement with
previous studies regarding the bar fraction as a function of morphological type,
in the sense that bars in early-type, low spinning galaxies, are longer than those in
late-type, high spinning ones (e.g. Erwin 2005).

An additional factor that might be playing an important role is
the content of cold gas.  Galaxies with high
spin have typically late-type morphology, with blue colors and higher
gas mass fractions. In these systems the ELN stability criterion might
not apply as pointed out by Christodoulou et al. (1995), given the 
big influence the gas has in the formation and development of stellar
bars. The threshold value found in simulations varies depending on
the amount of gas and the type of cooling that is implemented (Mayer \&
Wadsley 2004), but in general the gas component severely limits the
bar growth and evolution (Villa-Vargas et al. 2010), usually producing
weak, short-lived bars. Previously, 
Shlosman \& Noguchi (1993) studied the effect
of gas on the global stability of a galactic disk embedded in a live
halo, and they found two different regimes; in a low gas surface
density disk, the radial redistribution of the gas depends solely 
on the stability of the stellar disk, while for high gas surface
density, the gas develops inhomogeneities that heat up the stellar
component that increases the stability of the system preventing the
growth of a bar. This result fits well with our finding, where
long bars are hosted by luminous, massive, red galaxies that
typically present low gas content, while short bars are
found in low luminosity, low mass, blue galaxies which are
usually gas rich. 

On a recent study by Athanassoula et al. (2013), the authors follow the
formation and evolution of bars in N-body simulations of disk galaxies
with gas, where the gas component is modeled as a multiphase medium,
including star formation, feedback and cooling. The study shows that in gas-rich
simulations, the disk stays axisymmetric longer than in gas-poor ones,
and once the bar forms, it grows at a much slower rate. This explains why
fully grown bars are in place earlier in massive red disks
than in blue spirals.
In a subsequent study, we will investigate the
combined effect of gas content and spin on the bar properties
of disk galaxies.

\section{Conclusions}

In summary, we report a strong dependence of the bar fraction on the
galactic spin of disk galaxies. This dependence on $\lambda_{d}$ is different
for long and short bars. Long bars are preferentially found in galaxies
with low to intermediate $\lambda_{d}$ that are more
prone to develop bar instabilities
due to their self-gravitation. These galaxies are typically massive, luminous,
red, gas poor systems, when compared with galaxies hosting short bars. 
Instead, short bars are mostly found in high spinning galaxies, that at the
same time are typically low mass, faint galaxies, with blue colors and
rich in cold gas.

The rise and fall of the bar fraction for the case of long bars as a
function of $\lambda_{d}$ can be explained as the result of two
competing factors, the self-gravity of the systems and the support
by random motions. At high $\lambda_{d}$ values, the decline of the
bar fraction is due to the lack of self-gravity, being the disk more
sparse the global instabilities are suppressed or damped. At low
$\lambda_{d}$ values, the support of the system by random motions instead
of ordinate rotation becomes predominant; that also prevents the formation
and growth of bars.

Our finding of short bars being hosted by high spinning galaxies is
in good agreement with previous observational and theoretical studies,
considering that these systems are preponderantly blue, gas rich galaxies;
where the sparsity of the disk material and the large fraction of gas play
an important role on restricting the formation/growth of the bar.

\vspace{1cm}

The authors thank the referee for his/her constructive comments
that helped to improve the manuscript.
BC-S acknowledge E. Athanassoula and Issac Shlosman for helpful comments.
This work is supported by NSFC (no. 11173045 and 11233005), Shanghai
Pujiang Programme (no. 11PJ1411600) and the CAS/SAFEA 
International Partnership Programme for Creative Research Teams
(KJCX2-YW-T23)

Funding for  the SDSS and SDSS-II  has been provided by  the Alfred P.
Sloan Foundation, the Participating Institutions, the National Science
Foundation, the  U.S.  Department of Energy,  the National Aeronautics
and Space Administration, the  Japanese Monbukagakusho, the Max Planck
Society,  and the Higher  Education Funding  Council for  England. The
SDSS Web  Site is  http://www.sdss.org/.  The SDSS  is managed  by the
Astrophysical    Research    Consortium    for    the    Participating
Institutions. The  Participating Institutions are  the American Museum
of  Natural History,  Astrophysical Institute  Potsdam,  University of
Basel,  University  of  Cambridge,  Case Western  Reserve  University,
University of Chicago, Drexel  University, Fermilab, the Institute for
Advanced   Study,  the  Japan   Participation  Group,   Johns  Hopkins
University, the  Joint Institute  for Nuclear Astrophysics,  the Kavli
Institute  for   Particle  Astrophysics  and   Cosmology,  the  Korean
Scientist Group, the Chinese  Academy of Sciences (LAMOST), Los Alamos
National  Laboratory, the  Max-Planck-Institute for  Astronomy (MPIA),
the  Max-Planck-Institute  for Astrophysics  (MPA),  New Mexico  State
University,   Ohio  State   University,   University  of   Pittsburgh,
University  of  Portsmouth, Princeton  University,  the United  States
Naval Observatory, and the University of Washington.

\end{document}